\pdfoutput=1
\documentclass[pre,aps,twocolumn,showpacs,floatfix,nofootinbib,
superscriptaddress]{revtex4-1}

\usepackage{subfigure}
\usepackage{amssymb}
\usepackage{amsfonts}
\usepackage{amsmath}
\usepackage{amsthm}
\usepackage{epsfig}
\usepackage{graphicx}

\usepackage[usenames,dvipsnames]{color}
\usepackage[latin1]{inputenc}

\usepackage[hidelinks,unicode=true]{hyperref}
\hypersetup{colorlinks=true,
 	linkcolor=blue,
 	urlcolor=blue,
 	citecolor=blue,
 	pdfhighlight=/N
 }
 \usepackage{comment}
 \usepackage[normalem]{ulem}

\begin{document}

\title{Field emitter electrostatics: efficient improved simulation technique for highly precise calculation of field enhancement factors}

\author{Fernando F. Dall'Agnol}
\email{Author to whom correspondence on numerical issues should be addressed: fernando.dallagnol@ufsc.br}
\address{Department of Exact Sciences and Education (CEE), Universidade Federal de Santa Catarina,
 Campus Blumenau, Rua Jo\~{a}o Pessoa, 2514, Velha, Blumenau 89036-004, SC, Brazil}

\author{Thiago A. de Assis}
\email{thiagoaa@ufba.br}
\address{Instituto de F\'{\i}sica, Universidade Federal da Bahia,
  Campus Universit\'{a}rio da Federa\c c\~ao,
  Rua Bar\~{a}o de Jeremoabo s/n, 40170-115, Salvador, BA, Brazil}
\address{Instituto de F\'{\i}sica, Universidade Federal Fluminense,
  Avenida Litor\^{a}nea s/n, 24210-340, Niter\'{o}i, RJ, Brazil}

\author{Richard G. Forbes}
\email{Author to whom correspondence on general issues should be addressed: r.forbes@trinity.cantab.net}
\address{Advanced Technology Institute \& School of Computer Science and Electronic Engineering, University of Surrey, Guildford, Surrey GU2 7XH, UK}

\begin{abstract}

When solving the Laplace equation numerically via computer simulation, in order to determine the field values at the surface of a shape model that represents a field emitter, it is necessary to define a simulation box and, within this, a simulation domain. This domain must not be so small that the box boundaries have an undesirable influence on the predicted field values. A recent paper discussed the situation of cylindrically symmetric emitter models that stand on one of a pair of well-separated parallel plates. This geometry can be simulated by using two-dimensional domains. For a cylindrical simulation box, formulae have previously been presented that define the minimum domain dimensions (MDD) (height and radius) needed to evaluate the apex value of the field enhancement factor for this type of model, with an error-magnitude never larger than a ``tolerance" $\epsilon_{\rm{tol}}$. This MDD criterion helps to avoid inadvertent errors and oversized domains. The present article discusses (in greater depth than previously) a significant improvement in the MDD method; this improvement has been called the MDD Extrapolation Technique (MDDET). By carrying out two simulations with relatively small MDD values, it is possible to achieve a level of precision comparable with the results of carrying out a single simulation using a much larger simulation domain. For some simulations, this could result in significant savings of memory requirements and computing time. Following a brief restatement of the original MDD method, the MDDET method is illustrated by applying it to the hemiellipsoid-on-plane (HEP) and hemisphere-on-cylindrical-post (HCP) emitter shape models.


\end{abstract}

\maketitle


\section{Introduction}

Computer simulations are becoming a ubiquitous tool to investigate characterization parameters in field electron emission (FE) systems. In particular, numerical methods of analyzing the electrostatics of field emitters have been historically useful \cite{Edgcombe1,Edgcombe2,Edgcombe3,Forbes2003}, in that they have established some relatively simple formulas for parameters that characterize the electrostatic situation, often a dimensionless field enhancement factor (FEF).

In particular, electrostatic modeling of post-like shapes has attracted significant research interest. This modelling has usually been carried out: (a) using cylindrically symmetric classical-conductor post models; and (b) in so-called parallel-planar-plate (PPP) geometry \cite{ZhuRef,Podenok06,ZENG2009,Roveri16,Mauro2018,deAssisSchConj,deAssisJAP,FEF2021,APL2020,CPEE2017,DallAgnol2018,Screen3,Jen18,Forbes_preexponential}. This article also focuses mainly on cylindrically symmetric posts in PPP geometry.

In PPP geometry, a post of total height $h$ is assumed to stand on one of a pair of parallel planar plates of large lateral extent (very much greater than the physical \textit{separation} $d_{\rm{sep}}$ of the plates). The situation usually analyzed is the ``standard" one where (a) $h<<d_{\rm{sep}}$, and in addition (b) the \textit{gap-length} [$d_{\rm{gap}}=d_{\rm{sep}}-h$] between the post apex and the counter-electrode plate is very much greater than the post height [i.e., $h<<d_{\rm{gap}}$].       

This article also deals with this standard situation, but---due to the boundary conditions at the sides of the simulation box---the parallel plates are effectively of infinite extent.

However, these boundary conditions (discussed below) also generate mirror-image effects. These mirror-image effects lead to electrostatic depolarization of the emitter post being modelled, and hence to possible inaccuracy in the predicted FEF.

The larger the lateral dimensions of the simulation box, then the smaller will be the depolarization effects and the smaller will be the inaccuracy. A similar argument (but not quite the same) applies to the simulation-box height (see \cite{ESreview}). Large simulation boxes require additional computing resource, so there is scope for a trade-off between accuracy and box dimensions. This article is about optimising the trade-off and finding accurate FEF values.

At this point it will be useful to give some careful electrostatic definitions. A \textit{local electrostatic (ES) field} $E_{\rm{L}}$ is the ES field at some location ``L" on the post surface, typically a location relatively near the post apex. The local ES field at the post apex itself is denoted by $E_{\rm{a}}$.

A \textit{macroscopic ES field} is a field associated with the overall geometry of the system under discussion. There are several different kinds of macroscopic field (see \cite{ESreview}). In this article we are interested in the macroscopic ES field between the two parallel plates, in the absence of the post and in the absence of any effects that would be induced by its presence. For simplicity we refer to this inter-plate field as the \textit{plate field} and denote it by $E_{\rm{P}}$. 

In the same way that there are several different kinds of macroscopic field, there are several different kinds of dimensionless macroscopic field enhancement factor. The factor of interest here is the so-called local \textit{plate-field enhancement factor (PFEF)} $\gamma_{\rm{PL}}$ defined by
\begin{equation}
\gamma_{\rm{PL}} = E_{\rm{L}} / E_{\rm{P}} .
\label{gam-PL}
\end{equation}
Usually, it will be the apex PFEF value $\gamma_{\rm{Pa}}$, defined by
\begin{equation}
\gamma_{\rm{Pa}} = E_{\rm{a}} / E_{\rm{P}} ,
\label{gam-Pa}
\end{equation}
that is of particular interest.

Conventional classical electrostatic conventions are used here: this implies that for field electron emission the fields are negative but for field ion emission the fields are positive. FEF-values are, of course, positive in both cases.

Note that, as compared with some of our previous papers, our terminology and notation here have become more precise. The terms ``plate field" and ``plate FEF" (and related notation) have replaced the terms  ``macroscopic field" and ``macroscopic FEF" used in older papers, and the designation ``macroscopic" has now been given a wider meaning (see \cite{ESreview}).

Also note that we use the symbol $\gamma$ for a dimensionless FEF, rather than the symbol $\beta$ commonly used in FE literature. This is to avoid confusion with the alternative use of $\beta$ (and sometimes the alternaive use of the term ``field enhancement factor") to describe the relationship between field and some voltage-like quantity. All field enhancement factors discussed in this article are dimensionless, and we shall now stop using the term ``dimensionless".

The calculations described here use the \textit{Finite Elements Method (FEM)} \cite{Jin14,Galerkin} to solve Laplace's equation. This is done within the \textit{simulation domain} provided by a cylindrical simulation box with the post of interest on the box axis. The simulation domain is the volume (in 3D) or area (in 2D) between the post boundary and the box walls. Simulations are carried out using the COMSOL, v5.3, software package.

Initially, it has been and is necessary to consider posts for which there is an exactly known analytical expression $\gamma_{\rm{Pa}}^{\rm{(an)}}$ for the apex PFEF. The total percentage error $\epsilon_{\rm{num}}$ in the numerically derived (i.e., simulation-derived) apex-PFEF value, $\gamma_{\rm{Pa}}^{\rm{(num)}}$, is then defined as:

\begin{equation}
\epsilon_{\rm{num}} \equiv \left[ \frac{\gamma_{\rm{Pa}}^{\rm{(num)}}-\gamma_{\rm{Pa}}^{\rm{(an)}}}{\gamma_{\rm{Pa}}^{\rm{(an)}}} \right] \times 100 \%
\label{epsnum},
\end{equation}
This parameter $\epsilon_{\rm{num}}$ can in principle be positive or negative, but often our interest will actually be in the \textit{error magnitude} $\left|\epsilon_{\rm{num}} \right| $. Note that in some geometries, where no known analytical solution exists, it is necessary to replace the ``analytical" value by a numerically calculated ``highly precise" value. 

Also, note that the convention being used here differs in places from that used in our recent review \cite{ESreview}, and in these cases the change results in a change in the sign of $\epsilon_{\rm{num}}$. The revised definition here is more conventional, and thus may be clearer. 

The initial objective of the simulations was to determine \textit{minimum domain dimensions (MDD)}.This term is interpreted to mean the minimum height and radius of the simulation box that encloses the simulation domain, and the aim is to ensure that $|\epsilon_{\rm{num}}|$ will be less than some pre-specified \textit{error} $\epsilon_{\rm{tol}}$ in the FEF-value being determined. This parameter $\epsilon_{\rm{tol}}$ is always taken as positive, so strictly $\epsilon_{\rm{tol}}$ is an error magnitude. We call $\epsilon_{\rm{tol}}$ the simulation \textit{tolerance}.  

In the original paper \cite{JVSTB2019} on minimum domain dimensions by two of us, the chosen shape for the exact analytics was the ``hemisphere-on-plane" (HSP) model. However, in our recent review paper \cite{ESreview} we explored the more rewarding case of the ``hemiellipsoid-on-plane" (HEP) model.

The original paper \cite{JVSTB2019} also found an even more promising method of determining highly precise apex PFEF values, the so-called \textit{minimum domain dimensions extrapolation technique} (MDDET method). This method was subsequently used \cite{FEF2021} to determine precise apex-PFEF values for selected emitter shapes, and has been outlined in our review paper \cite{ESreview}.

The present article has two functions. It provides a brief review of our earlier work on the MDD and MDDET approaches. It also provides fuller discussions of these methods and of their application to the hemiellipsoid-on-plane shape model. It constitutes a fuller version of material presented at the 2022 International Vacuum Nanoelectronics Conference \cite{IVNC2022}.

We also need to make the point that this is an article about the electrostatics of classical conductors with smooth surfaces, and about how to implement this type of electrostatics effectively. There are also important issues, particularly with carbon nanotubes (e.g., \cite{JPCC2019}), about how to relate classical-conductor electrostatics to electrostatic treatments where atomic structure is taken into account. There have been important recent developments in this area, e.g. \cite{CJEHybrid1}. Related issues are outside the scope of the present paper, although we do think that considerations concerning minimum domain dimensions will play a part in any such wider discussions.

\section{The Minimum Domain Dimensions (MDD) method}

\begin{figure}
\includegraphics [scale=0.4] {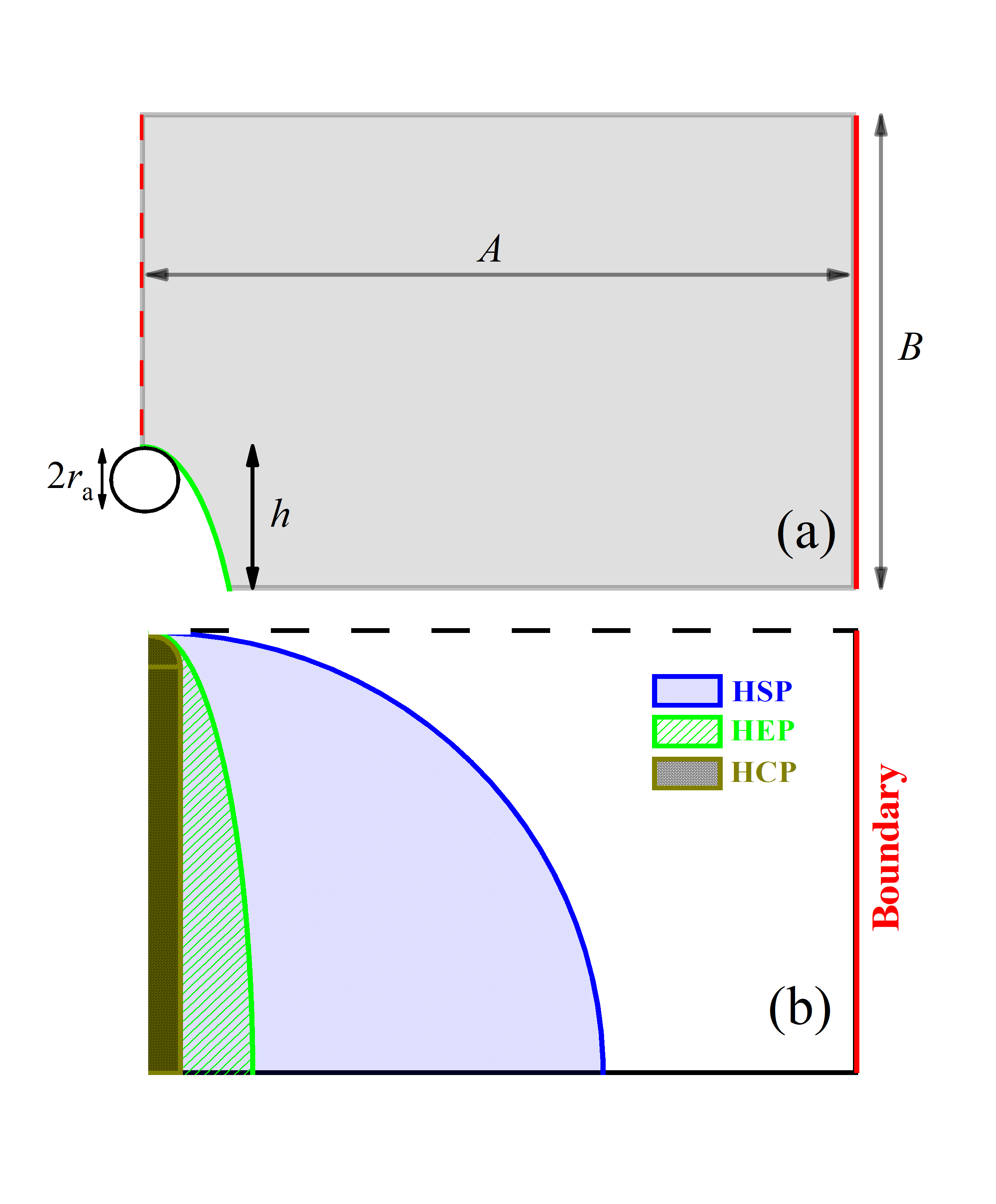}
\caption{(a) To show, in grey, the two-dimensional simulation domain $\Omega$. $A$ and $B$ are the characteristic dimensions to be minimized. The height $h$ and apex radius $r_{\rm a}$ of an emitter shape model are indicated. The interior of the model is removed because the emitter is considered a perfect conductor at Earth potential. (b) To show, for the hemisphere-on-plane (HSP), hemiellipsoid-on-plane (HEP) and hemisphere-on-cylindrical-post (HCP) shape models, each of which have the same apex sharpness ratio $h/r_{\rm{a}}$, the relative proximities of each shape to the lateral boundary.}
\label{Fieldsetup}
\end{figure}

Let us review the discussion of minimum domain dimensions \cite{JVSTB2019,ESreview}. As just indicated, the system considered was cylindrically symmetric with a post on axis. Profiles in a ``vertical" plane through the axis are shown in Fig. \ref{Fieldsetup}(a). The figure shows a half-plane with the original symmetry axis at the left-hand side. The domain width (i.e., radius) $A$ and height $B$ are the quantities that we aimed to minimize, for a given pre-specified tolerance  $\epsilon_{\rm{tol}}$.

We take the post to be perfectly conducting, which means no ES field penetration; hence, as already indicated, the post interior is not part of the simulation domain. Boundary conditions for the domain are imposed as follows. The post surface and the box bottom plane are set at Earth potential (taken as ES potential $\mathit{\Phi}$ = 0 V). At the right-hand side boundary, and at the left-hand-side boundary above the post, von Neumann-type boundary conditions are applied, with the specific requirement that the ES field component normal to the boundary is zero, at all points of the boundary. The top boundary is set as having a constant surface charge density $\sigma = \epsilon_0 E_{\rm{P}}$, where $\epsilon_{\rm 0}$ is the \textit{vacuum electric permittivity}.

Where relevant, we investigate the electrostatics properties of our system as a function of the \textit{apex sharpness ratio} $\sigma_{\rm a}\equiv h/r_{\rm a}$, where $r_{\rm a}$ is the \textit{apex radius of curvature}, and as a function of the domain dimensions $A$ and $B$.

Note the need to distinguish logically between the apex sharpness ratio, defined as above, and the \textit{aspect ratio}, which is defined as ``height"/``base-radius" ($\rho$), i.e. $h/ \rho$. For the hemisphere-on-cylindrical-post (HCP) model discussed below these two parameters have equal values, but this is not true for many shapes, including the HEP model.  For electrostatic field enhancement, the parameter of principal interest is the apex sharpness ratio.

\subsection{MDD for hemisphere-on-plane (HSP) model}

Formulae for the  minimum domain dimensions (MDD) were first  derived for a hemisphere-on-plane (HSP) emitter-shape model, and for that model can be written as follows \cite{JVSTB2019,FEF2021,ESreview}:

\begin{equation}
\left(\frac{A}{r_{\rm{H}}}\right)_{\rm{MDD-HSP}} = \frac{6}{\sqrt[3]{\epsilon_{\rm{tol}}}},
\label{mdd-hp-a/rh}
\end{equation}

\begin{equation}
\left(\frac{B}{r_{\rm{H}}}\right)_{\rm{MDD-HSP}} = \frac{5}{\sqrt[3]{\epsilon_{\rm{tol}}}}.
\label{mdd-hp-b/rh}
\end{equation}
Here, $r_{\rm{H}}$ is the radius of the hemisphere. The factors ``$6$" and ``$5$" are values rounded from $5.59$ and $4.64$, respectively \cite{JVSTB2019,FEF2021}. Rounding up the numerators makes the upper boundary for the MDD just a little larger, but makes the formulas easier to remember.

Obviously, in any particular case there is a need to decide a working tolerance value $\epsilon_{\rm{tol}}$. For exploratory numerical investigations we have typically taken $\epsilon_{\rm{tol}} = 1 \%$, as this keeps computation times low. For real-life applications one would normally want to choose a lower value. 

These hemisphere-on-plane results can in principle be used to decide MDD values for other post-like shapes, by inscribing them inside a hemisphere. For hemiellipsoid-on-plane (HEP) and hemisphere-on-cylindrical post (HCP) emitter-shape models, this process is illustrated in Figure \ref{Fieldsetup}(b). The boundary of the HSP shape is closer to the right-hand-side boundary of the simulation box than are the boundaries of the HEP and HCP shapes, so, the electrostatic influence of this box boundary will be larger on the HSP shape than it is on the other two shapes.

This can be demonstrated numerically, using $\epsilon_{\rm{tol}} =10 \%$ in eqs. (\ref{mdd-hp-a/rh}) and (\ref{mdd-hp-b/rh}). From the simulations, the numerical errors, as defined by eq.(\ref{epsnum}), are:  $\epsilon_{\rm{num}} =-9.4\%$ for the HSP shape;,  $\epsilon_{\rm{num}}=-1.83\%$ for the HEP shape with $\sigma_{\rm a}=30$; and $\epsilon_{\rm{num}}=-1.20\%$ for the HCP shape with $\sigma_{\rm a}=30$. The HSP and HEP results are found by using the analytical apex PFEFs. To determine $\epsilon_{\rm{num}}$ for the HCP model, we calculate $\gamma_{\rm{Pa}}^{\rm{(num)}}=27.3$ for $\epsilon_{\rm{tol}} = 10\%$ and  $\gamma_{\rm{Pa}}^{\rm{(num)}} = 27.632969$ for $\epsilon_{\rm{tol}} = 0.0001\%$, using the latter value of $\gamma_{\rm{Pa}}^{\rm{(num)}}$ as if it were an analytical value.

\begin{figure}
\includegraphics [scale=0.3] {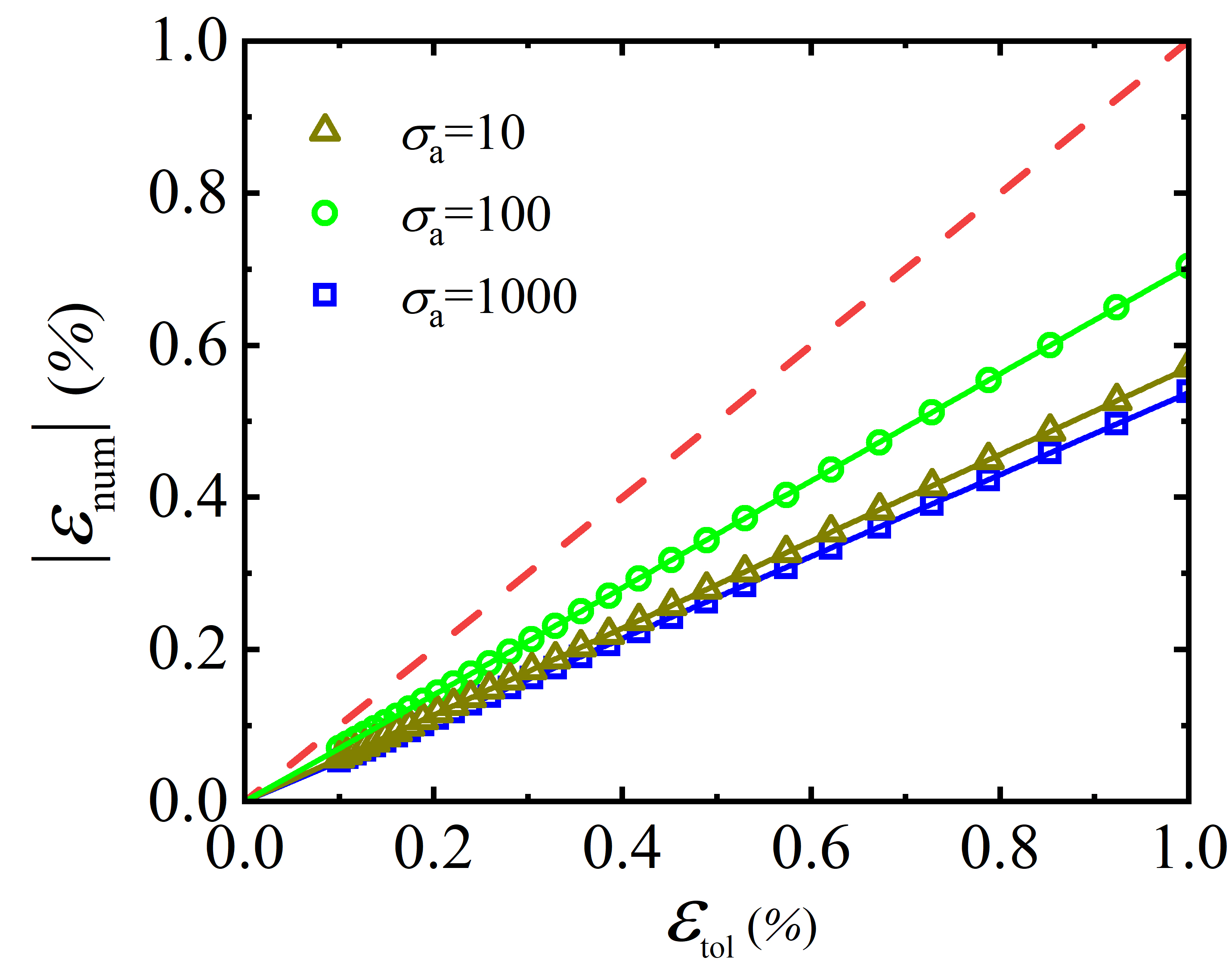}
\caption{To illustrate the results of applying the MDD method to hemiellipsoid-on-plane (HEP) emitter shape models, for apex-sharpness-ratio values $\sigma_{\rm a}=10$, $100$ and $1000$. The magnitude $|\epsilon_{\rm{num}}|$ of the actual numerical error applying to the relevant simulation is shown as a function of the specified ``tolerance" (maximum error-magnitude specified a-priori) $\epsilon_{\rm{tol}}$. In each case, linear behavior is observed. The dashed (red) line plots the relation $|\epsilon_{\rm{num}}| = \epsilon_{\rm{tol}}$; hence it can be clearly seen that $|\epsilon_{\rm{num}}|$ is always lower than $\epsilon_{\rm{tol}}$, as required. Filled lines are guides for the eyes. Note that, for given $\epsilon_{\rm{tol}}$,  $|\epsilon_{\rm{num}}|$ is not a monotonic function of $\sigma_{\rm a}$. Although not expected, we are not surprised by this; however, the precise reasons for the effect are not currently understood.}
\label{error2}
\end{figure}

In summary, the MDD that provide an error of magnitude not larger than $\epsilon_{\rm{tol}}$ for the HSP shape will ensure an error of magnitude smaller than this for the HEP shape inscribed in the HSP, and an even smaller-magnitude error for the HCP shape inscribed inside both.

\subsection{MDD for hemiellipsoid-on-plane (HEP) model}

This realization, and the availability of analytical formulae \cite{Lat81,Roh71,Kom91,Pog1, Pog2,Forbes2003} for the HEP model, led us to reformulate the MDD arguments on the basis of the theory of the HEP model. In this case the results depend on the apex sharpness ratio ($\sigma_{\rm{a}}$) of the HEP shape, and it was found empirically \cite{JVSTB2019,FEF2021,ESreview} that convenient formulas for minimum simulation-domain dimensions are
\begin{equation}
\left(\frac{A}{h}\right)_{\rm{MDD-HEP}} =
6 \times \sqrt[3]{\frac{f(\sigma_{\rm{a}})}{\epsilon_{\rm{tol}}}}
\label{mdd-hep2-A/h}
\end{equation}
\begin{equation}
\left(\frac{B}{h}\right)_{\rm{MDD-HEP}} =
5 \times \sqrt[3]{\frac{f(\sigma_{\rm{a}})}{\epsilon_{\rm{tol}}}},
\label{mdd-hep2-B/h}
\end{equation}
where
\begin{equation}
f(\sigma_{\rm{a}}) {= 0.2 + 0.8 \exp\left[-0.345 \left(\sigma_{\rm{a}}^{1/2} - 1\right)\right]}.
\label{f}
\end{equation}
Of course, when $\sigma_{\rm a}=1$, eqs. (\ref{mdd-hep2-A/h}) and (\ref{mdd-hep2-B/h}) reduce to eqs. (\ref{mdd-hp-a/rh}) and (\ref{mdd-hp-b/rh}), respectively.

Note that these formula do not work well if the chosen tolerance $\epsilon_{\rm{tol}}$ is so large that $A/h$ is close to unity or smaller. However, one would not normally want to work with a simulation box as small as this, so in practice this is not a significant constraint.

\subsection{MDD for hemisphere-on-cylindrical-post (HCP) model}

As with the HSP approach described earlier, formulae (\ref{mdd-hep2-A/h}) to (\ref{f})  can be used as upper bounds of the MDD for other shapes that can be inscribed in the HEP shape.

This is particularly useful when considering the widely used HCP shape model, for which no \textit{accurate} simple analytical formulae for the apex PFEF are known (probably none exist). The following exercise demonstrates this.

As before, by setting $\epsilon_{\rm{tol}}$ very low and thus using a relatively large simulation domain, we can make a reasonably accurate estimate of what the true apex PFEF is. This allows us to make good estimates of $\epsilon_{\rm{num}}$ for different values of $\epsilon_{\rm{tol}}$ and $\sigma_{\rm{a}}$ inserted into eqs. (\ref{mdd-hep2-A/h}) to (\ref{f}).   

Figure \ref{error2} shows the error-magnitude  $|\epsilon_{\rm{num}}|$ as a function of $\epsilon_{\rm{tol}}$, for $\sigma_{\rm a}=10$, $100$ and $1000$. Clearly, the results show that $|\epsilon_{\rm{num}}|$ is always lower than $\epsilon_{\rm{tol}}$. That is, the numerical error in the simulated result is always less than the specified tolerance.

\begin{figure}
\includegraphics [scale=0.3] {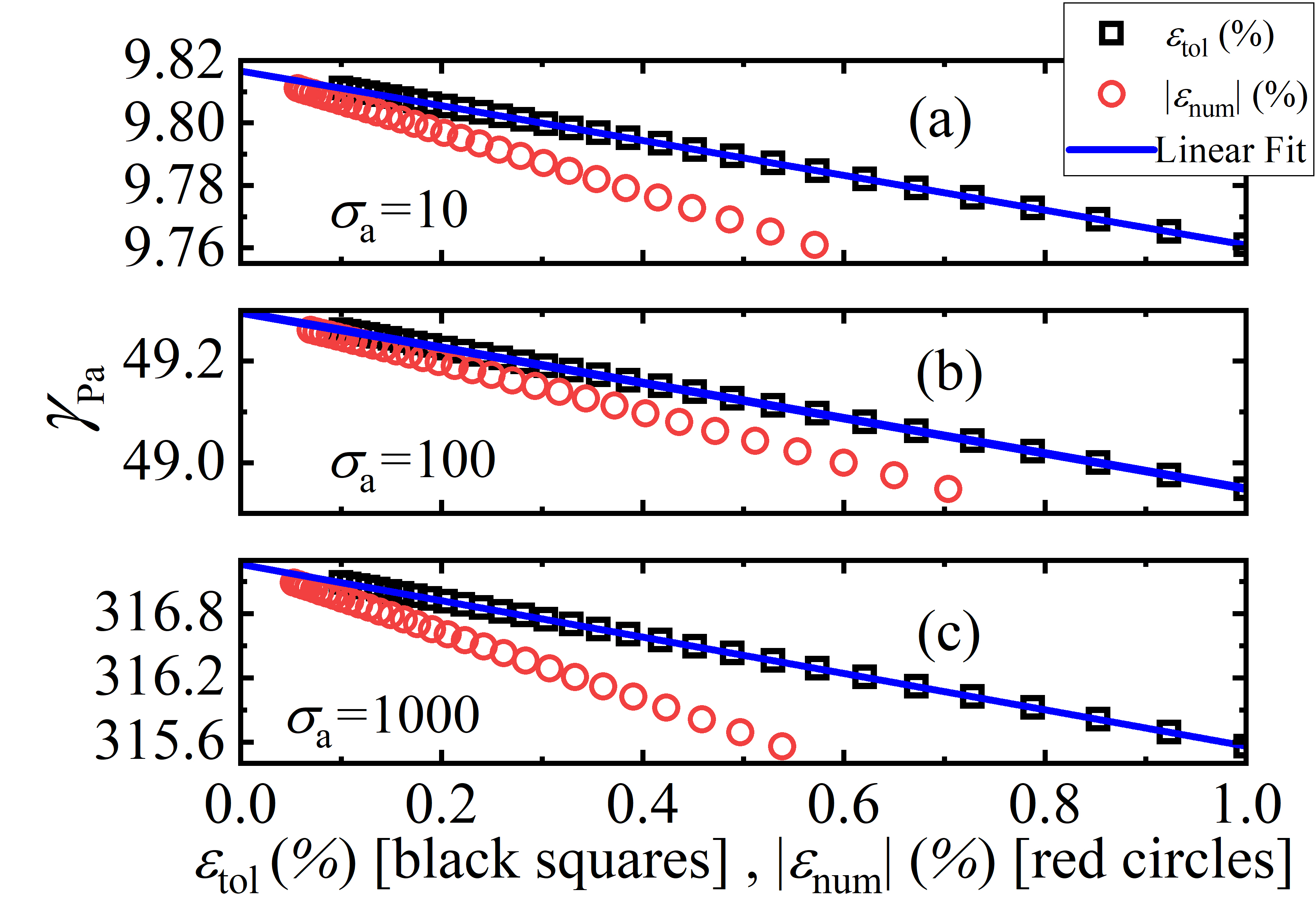}
\caption{To illustrate the results of applying the MDD Extrapolation Technique (MDDET method) to hemiellipsoid-on-plane (HEP) emitter shape models, for apex-sharpness-ratio values: (a) $\sigma_{\rm a}=10$; (b) $\sigma_{\rm a}=100$; (c) $\sigma_{\rm a}=1000$. In all cases, the (red) circles are the simulated apex PFEF ($\gamma_{\rm{Pa}}$) values, plotted as a function of the actual error magnitude, $|\epsilon_{\rm{num}}|$, which is found by comparison with exact analytical values. The (black) squares are these apex PFEF values plotted as a function of the specified input tolerance $\epsilon_{\rm{tol}}$. Since each red circle is to the left of the corresponding black square, this shows that (in all cases) the actual numerical error-magnitude is less than the specified tolerance.  Note the good linearity of the ``black-square plots".}
\label{FieldTermsnn}
\end{figure}

\bigskip

\section{the MDD Extrapolation Technique}

\subsection{MDDET method applied to the HEP shape model}

We now move on to consider the \textit{MDD Extrapolation Technique (MDDET)} for the HEP shape model. Calculations are carried out separately for HEP shape models  with apex sharpness ratios $\sigma_{\rm{a}}=10$, $100$ and $1000$. For each value of $\sigma_{\rm{a}}$, calculations are carried out for many values of the input specified tolerance $\epsilon_{\rm{tol}}$. These values lie in the range $0.1 \% \leqslant  \epsilon_{\rm{tol}} \leqslant 1 \% $, and are input into eqs (6) and (7).

Results are shown in Fig. \ref{FieldTermsnn}.  The sets of red circles show simulation-based apex-PFEF ($\gamma_{\rm{Pa}}$) values as a function of the numerical error-magnitude $|\epsilon_{\rm{num}}|$ derived from eq.(\ref{epsnum}).

The black squares show the same numerically derived apex-PFEF values, but as a function of the input specified tolerance $\epsilon_{\rm{tol}}$. In all cases it can be seen (because each red circle is to the left of the corresponding black square) that the actual numerical percentage error-magnitude is less than the specified tolerance. 

In all six cases these plots are essentially linear. If a regression line were fitted to any plot (each of which is based on 30 data points), then this line would intercept the vertical (``zero-error") axis at a value expected to be a good estimate of the true apex-PFEF value for the given input value of $\sigma_{\rm{a}}$.

In practice, we find that fitting regression lines to large sets of data-points is ``overkill". In practice, it seems to be adequate to estimate the intercept value by using two appropriately chosen points on (an extended version of) the ``black squares" line.  This ``two-point procedure" is more straightforward than fitting a regression line, and makes the MDDET method simple to implement.

The procedure we currently use for choosing two points on the ``black squares" line is as follows.

(i) Define the desired final tolerance $\epsilon_{\rm{tol}}$ for the MDDET result. This tolerance might typically be a value of order 0.001 \%.

(ii)  Define two much higher tolerance values $\epsilon_1 = 100 \; \epsilon_{\rm{tol}}$ and $\epsilon_2 = 1000 \; \epsilon_{\rm{tol}}$, which might typically become defined as 0.1 \% and 1 \%, respectively.  

(iii) Determine two apex-PFEF numerical estimates $\gamma_{\rm{Pa,1}}^{\rm{(num)}}$ and $\gamma_{\rm{Pa,2}}^{\rm{(num)}}$, using the sets of domain dimensions found by using $\epsilon_{1}$ and $\epsilon_{2}$ (respectively) as the inputs into eqs. (\ref{mdd-hep2-A/h}) and (\ref{mdd-hep2-B/h}).

(iv) Using these two data-points, an ``extracted estimate" $\gamma_{\rm{Pa}}^{\rm{(extr)}}$ of the apex PFEF can be obtained from the intercept on the vertical (``error = 0") axis, using the formula
\begin{equation}
\gamma_{\rm{Pa}}^{\rm{(extr)}} = \gamma_{\rm{Pa,1}}^{\rm{(num)}} 
- \epsilon_{1} \left[ \frac{\gamma_{\rm{Pa,2}}^{\rm{(num)}} - \gamma_{\rm{Pa,1}}^{\rm{(num)}}}{\epsilon_{2} - \epsilon_{1}}\right].  
\label{extrapol}
\end{equation}

It is not straightforward to make an \textit{empirically derived} estimate of the accuracy of this result. However, for the HEP model being discussed, the extracted apex-PFEF values can easily be compared with the exact analytical result. Some selected comparisons are made in Table \ref{tabI}.

\begin{table*}
 \centering
 \renewcommand{\arraystretch}{1.5}
 \caption{Apex plate field enhancement factors (PFEFs) and signed percentage errors for a hemiellipsoid-on-plane (HEP) emitter shape model, for $\epsilon_{\rm{tol}}=0.001$\%. Comparisons are made between the MDD and the MDDET methods, for several values of the apex sharpness ratio $\sigma_{\rm{a}}$. For the apex PFEFs, $\gamma_{\rm{Pa}}^{\rm{(an)}}$ is the exact analytical result, $\gamma_{\rm{Pa}}^{\rm{(num)}}$ is the numerical result from the MDD method, and  $\gamma_{\rm{Pa}}^{\rm{(extr)}}$ is the extracted result from the MDDET method. The related (signed) errors are denoted by $\epsilon_{\rm{num}}$ and $\epsilon_{\rm{extr}}$. The last column gives the ratio $ R_{\rm{mag}} $ of error magnitudes, as defined by $R_{\rm{mag}}=|\epsilon_{\rm{num}}|/|\epsilon_{\rm{extr}}|$.}

 \begin{tabular}{|c|c|c|c|c|c|c|}

  \hline
 
  $\sigma_{\rm{a}}(\rm{HEP})$ & $\gamma_{\rm{Pa}}^{\rm{(an)}}$ & $\gamma_{\rm{Pa}}^{(\rm{num})}$ & $\epsilon_{\rm{num}} (\%)$ & $\gamma_{\rm{Pa}}^{\rm{(extr)}}$ & $\epsilon_{\rm{extr}} (\%)$ & ratio \\ \hline
  - & - & (MDD) & (MDD) & (MDDET) & (MDDET) & $ R_{\rm{mag}} $ \\ \hline
 \hline

$10$   & $9.816638$ & $9.816587$ & $-0.000509$ & $9.816642$ & $+0.000047$ & $11$   \\ \hline
$20$   & $15.386358$ & $15.386284$ & $-0.000483$ & $15.386371$ & $+0.000085$ & $5.7$  \\ \hline
$50$   & $29.324367$ & $29.324211$ & $-0.000529$ & $29.3244412$ & $+0.000253$ & $2.1$  \\ \hline
$100$  & $49.295371$ & $49.295096$ & $-0.000558$ & $49.295574$ & $+0.000412$ & $1.4$  \\ \hline
$200$  & $84.701121$ & $84.700650$ & $-0.000556$ & $84.701546$ & $+0.000501$ & $1.1$  \\ \hline
$500$  & $177.975511$ & $177.974684$ & $-0.000464$ & $177.976428$ & $+0.000515$ & $0.9$  \\ \hline

\end{tabular}
\label{tabI}
\end{table*}

\begin{table*}
 \centering
 \renewcommand{\arraystretch}{1.5}
 \caption{Apex PFEFs and signed percentage errors for a hemisphere-on-cylindrical-post (HCP) emitter shape model, for $\epsilon_{\rm{tol}}=0.001\%$. Other notations are as described in the caption to Table \ref{tabI}, except that $\gamma_{\rm{Pa}}^{\rm{(precise)}}$ replaces $\gamma_{\rm{Pa}}^{\rm{(an)}}$.}

 \begin{tabular}{|c|c|c|c|c|c|c|}

  \hline
 
   $\sigma_{\rm{a}}(\rm{HCP})$ & $\gamma_{\rm{Pa}}^{\rm{(precise)}}$ & $\gamma_{\rm{Pa}}^{(\rm{num})}$ & $\epsilon_{\rm{num}} (\%)$ & $\gamma_{\rm{Pa}}^{\rm{(extr)}}$ & $\epsilon_{\rm{extr}} (\%)$ & ratio  \\ \hline
  - & - & (MDD) & (MDD) & (MDDET) & (MDDET) & $ R_{\rm{mag}} $ \\ \hline
 \hline

$10$   & $11.795696$ & $11.795792$ & $+0.000809$ & $11.795791$ & $+0.000805$ & $1.0$   \\ \hline
$20$   & $19.986730$ & $19.986665$ & $-0.000322$ & $19.986759$ & $+0.000143$ & $2.2$   \\ \hline
$50$   & $42.205122$ & $42.205149$ & $0.000064$ & $42.205087$ & $-0.000082$ & $0.78$   \\ \hline
$100$  & $76.331892$ & $76.331393$ & $-0.000653$ & $76.331757$ & $-0.000176$ & $3.7$   \\ \hline
$200$  & $140.359346$ & $140.359108$ & $-0.000169$ & $140.360266$ & $+0.000656$ & $0.26$  \\ \hline
$500$  & $319.766109$ & $319.765436$ & $-0.000210$ &  $319.767827$ & $+0.000537$ & $0.39$ \\ \hline
$1000$ & $602.263731$ & $602.263189$ & $-0.000090$ &  $602.266315$ & $+0.000428$ & $0.21$ \\ \hline

\end{tabular}
\label{tabII}
\end{table*}

Table \ref{tabI} makes comparisons of apex PFEF values, for six values of the apex sharpness ratio $\sigma_{\rm{a}}$. Column 2 shows values of the exact analytical result, which corresponds to the formal situation where the simulation box dimensions become infinitely large. Columns 3 and 4 show the apex-PFEF values, and corresponding percentage errors (as compared with the analytical result), derived via the MDD method. Columns 4 and 5 show equivalent data for the MDDET method. In all cases the underlying required tolerance has been taken as $\epsilon_{\rm{tol}} = 0.001$ \%. Finally, column 6 shows the ratio $\epsilon_{\rm{num}}/\epsilon_{\rm{extr}}$.

Note that in all cases, for both the MDD and MDDET methods, the magnitude of the numerical error in these methods is less than the specified tolerance.

For apex sharpness ratios less than around 200, the MDDET method gives smaller numerical-error magnitudes than does the MDD method, as shown in the last column of Table \ref{tabI}.

The real advantage of the MDDET method is the saving on computer memory requirements and on computing time, as compared with the MDD method. This is because it is quicker to carry out two high-tolerance calculations (using small simulation-box dimensions) than to carry out one low-tolerance calculation (using large simulation-box dimensions).

\begin{figure}
\includegraphics [scale=0.3] {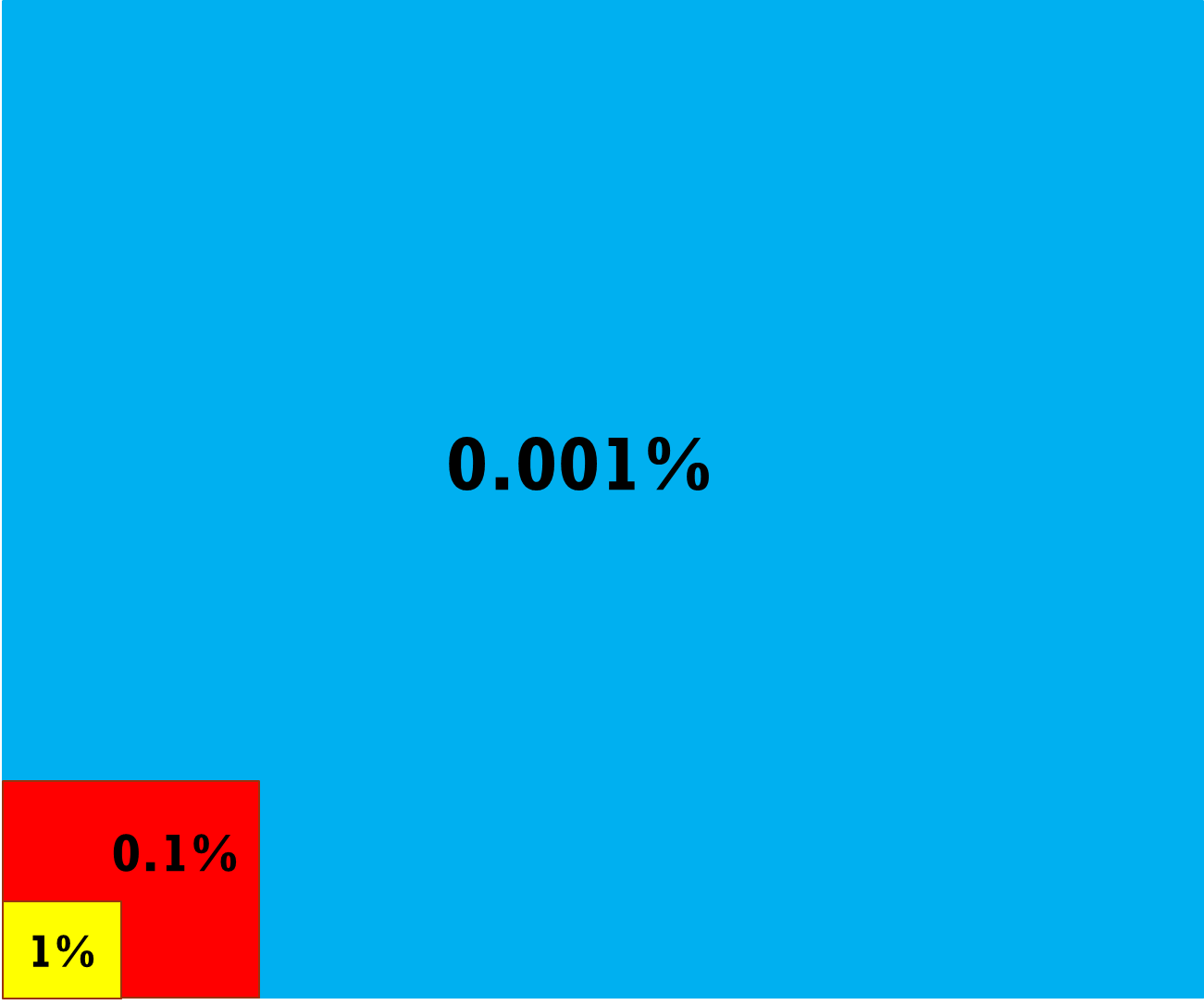}
\caption{Comparison of the predicted areas of the two-dimensional simulation domains needed to guarantee a tolerance (i.e., error-magnitude upper bound) of the percentage error level shown. The MDDET method uses two simulations based on the two smaller simulation domains, rather than a single calculation based on the large (blue) simulation domain.  This can lead to savings in memory requirements and computing time.} 
\label{DomainsN}
\end{figure}

Figure \ref{DomainsN} illustrates the basic principle of how this saving arises. The coloured rectangles are visual representations of the sizes of the relevant simulation domains. The two small rectangles combined have area approximately $1/20$th and volume $1/90$th of the large rectangle. The memory requirements and computer times ``go with" the sizes of the rectangles, though not in any straightforward fashion. The relatively small volume associated with the two small rectangles combined makes the MDDET method especially advantageous in 3D models.

Another context in which the MDDET method has an advantage can be illustrated as follows. If, for some technical computing reason, the maximum domain size that can be implemented corresponds to the red (0.1 \%) square in Fig. \ref{DomainsN}, then the apex-PFEF estimate provided by the MDDET method can be much more precise than that provided by the MDD method.

\subsection{MDDET method applied to the HCP shape model}

We now discuss applying the MDDET method to the hemisphere-on-cylindrical-post (HCP) shape mode. In this case, there is no known analytical solution, so we use instead the result $\gamma_{\rm{Pa}}^{\rm{(precise)}}$ of a highly precise numerical estimate, made using a tolerance of 0.001 \%. Otherwise, procedures are similar to those described in the previous section. Results are shown in Table \ref{tabII}, in the same form as in Table \ref{tabI}.

In this HCP-model case there is often no significant gain in numerical accuracy when the MDDET method is used (if anything the reverse). Rather, the advantage (in principle) of the MDDET method is the reduction in memory requirements and computing time. For any given simulation geometry, there can be a separate question of whether these advantages are significant enough to be practically useful.

The advantage of the MDDET over the MDD regarding computational time and memory can be checked directly by simple comparison. In Table \ref{tabIII}, the time on the 2nd column is the combined computational time to perform the MDDET with tolerance 0.001 \%. The 3rd column is the time needed to perform the ordinary MDD with same tolerance of 0.001 \%. For simulations that spans several parameters, saving time is of the essence. However, this advantage is not so straightforward to quantify.

The memory used by the simulator tends to increase with the number of mesh elements (i.e. with the apex sharpness ratio $\sigma_{\rm{a}}$). However, the dependency is somewhat irregular. Further, since the software packagae COMSOL is not open source, it is not possible to access the way that its memory allocation management works; this, it is difficult to understand exactly why the memory does not scale with the number of elements. 

\begin{table*}
 \centering
 \renewcommand{\arraystretch}{1.5}
 \caption{Memory resources}

 \begin{tabular}{|c|c|c|c|c|}

  \hline
 
$\sigma_{\rm{a}}(\rm{HCP})$ & time (s) & time (s) & memory (GB) & memory (GB) \\ \hline
& MDDET & MDD & MDDET & MDD \\ \hline

$10$  & $89$ & $445$ & $\approx 1.5 $ & 2.2 \\ \hline
$50$  & $118$ & $1540$ & $1.6$ & $1.8$ \\ \hline
$100$ & $188$ & $2242$ & $3$ & $7.8$   \\ \hline
$200$ & $265$ & $4360$ & $5.1$ & $17$ \\ \hline

\end{tabular}
\label{tabIII}
\end{table*}

The number of mesh elements used in a simulation is more than sufficient for the actual numerical error-magnitude to beat the specified tolerance. As has been discussed, no matter how many mesh elements there are, the precision of the simulastion will always be hindered by the proximity of the boundaries.

\subsection{Methods for other shape models}

To apply either the MDD or the MDDET method to emitter-shape models other than the HCP, one inscribes the given shape into a hemi-ellipsoid. This is done for the HCP model shape in Fig. \ref{Fieldsetup}. The apex sharpness ratio $\sigma_{\rm a}$ of the enclosing HEP-model can then be used as an upper limit for the model-shape of interest.

\section{Comments and Summary}

\subsection{Choice between MDD and MDDET methods}

In general, if only limited precision is needed for apex or other PFEF values (say, $\epsilon_{\rm{tol}} \geq 0.1 \%$), then we recommend using the MDD method (rather than the MDDET method). The MDD method is  simpler, because the simulation only needs to be carried out once.

In fact, if we try to apply the MDDET method for low values of precision (corresponding, say, to high tolerance values $\epsilon_{\rm{tol}} > 10\%$) then the method may not work well.  The MDDET method is particularly advantageous, in terms of practicality and reliability, if the precision requirement is exceptionally high, say, $\epsilon_{\rm{tol}} \leq 0.001\%$.

\subsection{Extension to infinite rectangular arrays}

The arguments in this paper can be readily  extended to the analysis of infinite square and rectangular arrays. Such arrays are modelled by placing the post at the centre of the base of a square or rectangular simulation box. The mirror images in the side-walls, together with the central post, constitute the array, with the spacings in the array determined by the lengths of the sides of the box base.

Nearly all part of the walls of the square or rectangular box would be farther from the post than the walls of a cylindrical simulation box with radius equal to half the length of the shorter side of the base of the square of rectangular box. Hence, for a specified tolerance, the precision of PFEF results for the array would be expected to be better than precision of PFEF results for the singe-post case.

\subsection{Meshing errors}

So far, the whole discussion has been about the influence of the simulation box boundaries (and hence the role of the domain dimensions) on the error-magnitude $\epsilon_{\rm{num}}$ of a simulated field-enhancement-factor value.  However, there are other potential sources of error in the system, with the chief of these being the way that the finite-element mesh is chosen and refined. If the resulting mesh is ``sufficiently fine", particularly near the apex of the chosen shape, then the meshing-error-magnitude can be made significantly less than the simulation-error-magnitude due to the proximity of the boundaries to the central post-like shape, and the former can then be disregarded. Our practice is to always work in the regime where meshing errors are so small in magnitude that they can be disregarded.

As a normal working starting point, we assume that, near the shape apex, the mesh length should be of order $r_{\rm{a}}/30$ or less, where $r_{\rm{a}}$ is the apex radius of curvature. When a simulated FEF-value has been achieved, we check that changing the meshing does not significantly alter the derived FEF-value, at the level of significance of interest. If this is not the case, then the meshing is refined until effective independence of the simulated result from the meshing details is achieved, causing the simulated result to be ``safe from meshing errors". All results in this paper are thought to be safe from meshing errors.

\subsection{Summary}

A summary of our conclusions is as follows.  The MDD and MDDET methods have been described previously and have been restated here. The usefulness of the MDD method is that it allows researchers to minimise the dimensions of a simulation domain, in order to obtain a result of specified accuracy. This provides benefits in terms of computational time and memory required.

More important, we have provided a better formal structure for the MDDET method. The essential advantage of the MDDET method is that, for a given required tolerance, it allows the use of much smaller simulation domains; this is particularly useful in complex time-consuming computations. For apex (or other) PFEF values, the MDDET method allows a result of specified accuracy (or better) to be obtained by carrying out two calculations using simulation boxes of small dimensions, rather than one box of large dimensions. This approach can be used for any cylindrically symmetric shape that can be inscribed within a hemiellipsoid of revolution. 

However, if the precision requirement is low (say, $\epsilon_{\rm{tol}}\geq0.1\%$), then using the MDD method may be simpler and preferable.

\section{Acknowledgement}

T A dA is grateful for funding from the Conselho Nacional de Desenvolvimento Cient\'{i}fico e Tecnol\'{o}gico (CNPq) (No. 310311/2020-9) and the Funda\c{c}\~{a}o de Amparo \`{a} Pesquisa do Estado do Rio de Janeiro (FAPERJ) (No. E-26/203.860/2022).

\section{Author Declarations}

\subsection{Conflicts of Interest}

The authors have no conflicts to disclose.

\subsection{Data availability}

The data that supports the findings of this study are available within the
article.

\subsection{Authors' contributions}

\textbf{Fernando Dall'Agnol}: conceptualization (supporting); data curation (equal); formal analysis (equal); methodology (equal); project administration (equal); resources (equal); software (equal);supervision (equal); validation (equal); visualization (equal); writing/original draft (lead); writing/reviewing and editing (supporting).

\textbf{Thiago de Assis}: conceptualization (lead); data curation (equal); formal analysis (equal); funding acquisition (lead); methodology (equal); project administration (equal); resources (equal); software (equal); supervision (equal); validation (equal); visualization (equal); writing/original draft (supporting); writing/reviewing and editing (supporting).

\textbf{Richard Forbes}: conceptualization (supporting); data curation (equal); formal analysis (equal); methodology (equal); project administration (equal); resources (lead); supervision (equal); validation (supporting); visualization (equal); writing/original draft (supporting); writing/reviewing and editing (lead).

\bibliography{Bibliography2}

\end{document}